\documentclass[prc,twocolumn,showpacs,aps]{revtex4}
\usepackage{epsfig,graphicx,float,pst-all, rotating}
\usepackage[utf8x]{inputenc}
\usepackage{hyperref}
\usepackage{amssymb}
\usepackage{amsmath}
\usepackage{multirow}

\newcommand{\be}{\begin{equation}} 
\newcommand{\ee}{\end{equation}}
\newcommand{\bea}{\begin{eqnarray}} 
\newcommand{\eea}{\end{eqnarray}}
\newcommand{\bc}{\begin{center}}
\newcommand{\ec}{\end{center}}

\begin{document}

\title{Nuclear fusion as a probe for octupole deformation in $^{224}$Ra} 

\author{Raj Kumar}
\email{rajkumarfzr@gmail.com}%

\author{J. A. Lay}
\email{lay@pd.infn.it}

\author{A. Vitturi}

\affiliation{ Dipartimento di Fisica e Astronomia ``Galileo Galilei", Universit\`{a} di Padova, via Marzolo, 8, I-35131 Padova, Italy}
\affiliation{INFN, Sezione di Padova, via Marzolo, 8, I-35131 Padova, Italy}

\begin{abstract}
\begin{description}
\item[Background] Nuclear fusion has been shown to be a useful probe to study the different nuclear shapes. However, the possibility of testing octupole deformation of a nucleus with this tool has not been fully explored yet. The presence of a stactic octupole deformation in nuclei will enhanced a possible permanent electric dipole moment, leading to a possible demonstration of parity violation.
\item[Purpose] To check whether static octupole deformation or octupole vibration in fusion give different results so that both situations could be experimentally disentangled.
\item[Method] Fusion cross sections are computed in the Coupled-Channels formalism making use of the Ingoing-Wave Boundary Conditions (IWBC) for the systems $^{16}$O+$^{144}$Ba and $^{16}$O+$^{224}$Ra.
\item[Results] Barrier distributions of the two considered schemes show different patterns. For the $^{224}$Ra case, larger differences are found in correspondence with its larger octupole deformation.
\item[Conclusions] The measurement of barrier distributions can be a complementary probe to clarify the presence of octupole deformation. 
\end{description}
\end{abstract}

\maketitle


\section{Introduction}

The search of octupole deformations in nuclei is living a revival thanks to its impact in a possible permanent atomic electric dipole moment (EDM)~\cite{Nature,Gri09}. A non-zero EDM will indicate a time-reversal (or equivalently charge-parity) violation. The magnitude or the experimental maximum limit to it can constraint the different suggested extensions to the standard model~\cite{Pos05}. The presence of a static octupole deformation in an odd nucleus will generate enhanced nuclear Schiff moments, which  contributes to the atomic EDM so that it can be improved by several orders of magnitude~\cite{Dob05}.

Therefore, the experimental focus is set on the search of permanent octupole deformations in some of the regions where different theoretical approaches predict strong octupole correlations~\cite{But96,Lis13} with the help of continuous development of the radioactive beam facilities. In the present work we will focus on $^{144}$Ba and $^{224}$Ra as representative of two different regions where possible static deformations have been predicted.

In~\cite{Nature,Wol93}, Coulomb excitation has been used to measure the different electric transition probabilities. This tool provides quadrupole and octupole transition probabilities with good accuracy. Smaller dipole electric transition probabilities will carry larger error bars, even though this problem is experimentally affordable and they were able to provide some measurements. It should be kept in mind that large octupole transitions can be found also for dynamic octupole vibrations. In addition, the coupling between quadrupole deformation and octupole vibrations can lead to enhanced dipole moments~\cite{But96}.

Therefore, we would like to propose here a complementary experimental probe for static octupole deformation. A traditional experimental tool for the study of nuclear structure is provided by subbarrier fusion. It is well known that fusion at energies around the Coulomb barrier is driven by the dynamical couplings to the internal degrees of freedom of the fusing counterparts~\cite{Bal98,Hag12}. The absence or presence of octupole and dipole moments in one of the fusing partners will have a certain impact in the final subbarrier fusion cross section. This fact could suggest, if the minimum accuracy is reached, the possibility of distinguishing between static octupole deformation and the corresponding dynamical vibration.

The work is structured as follows. In Sec.~\ref{theo} we recall the reaction formalism for studying nuclear fusion including deformations and/or vibrations. We apply this framework in Sec.~\ref{appli} to the reactions $^{16}$O+$^{144}$Ba and $^{16}$O+$^{224}$Ra considering quadrupole deformations for $^{144}$Ba and $^{224}$Ra looking for the differences between adding an octupole vibration or deformation to the previous quadrupole deformation. Finally, in Sec.~\ref{sum} the main results
of this work are summarized.

\section{\label{theo} Reaction framework}

Fusion probabilities are calculated by solving the corresponding coupled-channel equations under ingoing-wave boundary conditions (IWBC). The coupled-channel formalism for direct reaction processes given by Austern~\cite{Aus87} expands the total wave function in terms of the wavefunction for the internal state of the projectile $\phi_{\beta}$ and the radial wave functions $\chi_{\beta}$ that accounts for the relative motion between projectile and target. 
This leads to a set of coupled equations for the radial wave functions:
\begin{eqnarray}
\frac{d^{2}\chi_{\beta}}{dR^{2}}-\frac{L(L+1)}{R^{2}}+\frac{2\mu}{\hbar^{2}}[E_{\beta}-V_{\beta}^{eff}(R)]\chi_{\beta} = \nonumber \\
=\frac{2\mu}{\hbar^{2}}\sum_{\alpha\ne\beta}V_{\beta\alpha}^{coup}(R)\chi_{\alpha}  
\label{eq.3}
\end{eqnarray}
In these expression 
$V$ is the interaction potential, 
$\mu$ is the reduced mass, 
and $E_{\beta}$ is the relative energy. 
Each channel $\beta$ corresponds to a set of quantum numbers \{$I,L,J$\}, where $I$ is the angular momentum of the internal state, 
 $L$ is the angular momentum in the relative coordinate, and both are coupled to a total angular momentum $J$.

As a simplification we can use the same central potential for the $V_{\beta}^{eff}(R)$ although this potential is modified by the terms $V_{\beta\beta}$ known as reorientation terms. The non central terms $V_{\beta\alpha}$ are those on charge of the coupling between two different channels $\alpha$ and $\beta$. 
For all the cases, we will consider, for the central 
potential, a Woods-Saxon 
with the parametrization from Aky\"{u}z-Winther~\cite{Aky81b,BW} plus the corresponding Coulomb repulsion.

In our case we will consider one of the two ions to be a quadrupole deformed rotor.
 We will also look at cases where the deformed nuclei has an octupole vibration or a static octupole deformation together with the quadrupole one.

In the case of permanent deformation, 
 we can describe the radius of the deformed nucleus as a function of the angle $\theta'$, defined with respect to an intrinsic \textit{body-fixed} frame, $R(\theta') = R_0[1  + \sum_{\lambda} {\beta}_{\lambda} Y_{\lambda 0}^*(\theta')]$,
where $R_0$ is an average radius of the deformed nucleus and $\beta_{\lambda}$ a dimensionless deformation parameter. 
%

\begin{table}
\caption{\label{Tab:betas} Theoretical deformation parameters for $^{144}$Ba and $^{224}$Ra according to~\cite{Naz84} for the different multipolarities. }
\begin{center}
\begin{tabular}{ccc}
\hline
 & $\quad$ $^{144}$Ba $\quad$ & $\quad$ $^{224}$Ra $\quad$\\
\hline  
\hline
$\beta_{2}$  &  $0.149$  &   $0.138$   \\  
\hline
$\beta_{3}$  &  $0.068$   &  $0.099$   \\
\hline

\end{tabular}
\end{center}
\end{table}

The nuclei of interest in the present work are candidates to have a permanent octupole deformation together with the quadrupole one. In Table~\ref{Tab:betas}  we show the deformation parameters used in this work following the theoretical predictions from~\cite{Naz84}. 


If one assumes that the potential is still a function of the distance between projectile and target, the potential can be expanded in multipoles as:
\be
V(R,\Omega)= \sum_{\lambda \mu } V_{\lambda}(R) {\cal D}^{\lambda}_{\mu 0}(\alpha,\beta,\gamma) Y^{*}_{\lambda \mu }(\hat R),
\label{1}
\ee
where ${\cal D}$ is the so called {\textit rotation matrix} (or $D$-matrix)~\cite{BM}. 
 Finally, evaluating the matrix elements of this potential between the states of the rotor, it is possible to obtain the coupling potentials. See~\cite{BM,Tam65,Bal98,Hag12} for more details.


%
%

The spherical harmonic will connect two states $\chi_{\alpha}(R)$ and $\chi_{\beta}(R)$ depending on the order $\lambda$. The iso-centrifugal approximation~\cite{Tan87,Gom86,Hag04} is used to reduce the size of the calculation.

On the other hand, the coupling may also arise from a vibration of one of the nucleus. In this case this vibration is characterized as a variation on the surface as $R(\xi)=R_0 [ 1  + \sum_{\lambda, \mu} \alpha_{\lambda \mu} Y^{*}_{\lambda \mu}(\hat{R}) ]$,
where $\alpha_{\lambda \mu}$ are to be understood as dynamical variables, given in terms of phonon creation 
($b^{\dag}_{\lambda \mu}$) and annihilation ($b_{\lambda \mu}$) operators~\cite{Bro78}.  
%
%
The nuclear coupling between the ground state and the first one phonon state of multipolarity $\lambda$ reads:
\begin{equation}
V_{coup}=-\frac{\beta_\lambda}{\hat{\lambda}}R_{0}\frac{\partial V}{\partial R}Y^{*}_{\lambda \mu}(\hat{R}).
\end{equation}
%
However, we will consider an octupole vibration on top of a quadrupole deformed nucleus instead of a pure spherical nucleus. In this case, the derivative of the potential has a certain dependence of the orientation. We can express such dependence in the form:
\begin{equation}
V_{coup}(R,\xi) = - \frac{\beta_3}{\hat{3}} R_{0}\frac{\partial V}{\partial R}\left( R-R_{0}\left[1+\beta_{2}Y_{20}\right]\right)Y_{30} ,
\label{vibformfactor}
\end{equation}
which we later expand in spherical harmonics as done for the full rotor case, i.e. following Eq. \eqref{1}. Here, we have selected $\mu=0$ because the coupling of the octupole vibration with the quadrupole deformation splits the strength into different bands, being $K=0^{-}$ the lowest one~\cite{Don66,Alo95}. In other words, if the nuclei is significantly prolate, the vibration is stronger along the symmetry axis where the radius is larger.  
One should remind that this potential is only coupling zero octupole phonon states with one octupole phonon states. Together with these vibrational couplings, the traditional quadrupole deformed potential will only act between states with the same number of phonons.

%
%


%


In addition to all these nuclear couplings considered here, high order Coulomb couplings should be considered. However, it has been shown that Coulomb couplings has a minor role compare to the nuclear ones~\cite{Hag97} and, therefore, they will not be included here.

Finally, solving the coupled equations a probability of transmission $T_{L}$ for each angular momentum $L$ is obtained. The fusion cross-section 
is given by
\begin{equation}
\sigma=\sum_{L=0}^{L_{max}}\sigma_{L}=\frac{\pi\hbar^2}{2\mu E}\sum_{L=0}^{L_{max}}(2L+1)T_{L}(E).
\label{eq:14}
\end{equation}
The probability of transmission for each partial wave can also be calculated simply by a shift of energy,
\begin{equation}
T_{L}\cong T_{0}\left[E-\frac{L(L+1)\hbar^{2}}{2\mu R_{0}^2} \right],
\label{eq:15}
\end{equation}
where $R_{0}$ is the position of the barrier for the s-wave~\cite{Bal98}. Hereafter we will refer as transmission probability $T$ to the probability of transmission for the s-wave $T_{0}$~\cite{Bal98,Kum14}.
The derivative of the transmission probability, 
usually called barrier distribution, 
is approximately proportional to the second derivative of the product of the cross section and the energy ($E\sigma$), thus being a direct link between nuclear structure and this experimental observable~\cite{Row91,Bal98}.

\section{\label{appli}Applications}

\subsection{Prolate vs. Oblate deformations}

\begin{figure}
\includegraphics[width=0.9\columnwidth]{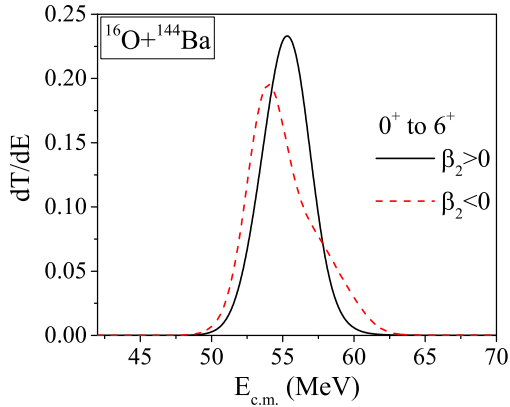}
\caption{(Color online) Barrier distribution for the fusion of $^{16}$O and $^{144}$Ba including a positive quadrupole deformation for $^{144}$Ba, solid line, and the same for a negative quadruople deformation, dashed line. 
}
\label{figxsprovsobl}
\end{figure}

Fusion reaction is one of the experimental tests that can discriminate between prolate and oblate deformations. 
A change on the sign of the deformation length will change the sign of the coupling and reorientation terms. 
 The sign of the off-diagonal coupling potential does not have an effect in the cross section. 
 Instead, the reorientation term is on-diagonal, so being the main responsible for the change in the cross section when one of the counter partners is either oblate or prolate.





In Fig.~\ref{figxsprovsobl} we show the barrier distribution for the fusion of the system $^{16}$O+$^{144}$Ba considering a positive quadrupole deformation parameter, solid line, and a negative one, dashed line. In this calculation and hereafer, we consider only the first three leves of a typical quadrupole deformed rotor, including all allowed couplings. The influence of higher energy and spin levels depends on the nucleus. For $^{144}$Ba we have studied the variation of the barrier distribution with the number of levels included. 
From this analysis we concluded 
that including more levels does not alter dramatically the barrier distribution. Similar results are found for $^{224}$Ra.

Additional examples of fusion reactions with prolate and oblate nuclei can be found in~\cite{Bie96,Bal98}

\subsection{Coupled-Channels vs. frozen approximation}

\begin{figure}
\includegraphics[width=0.9\columnwidth]{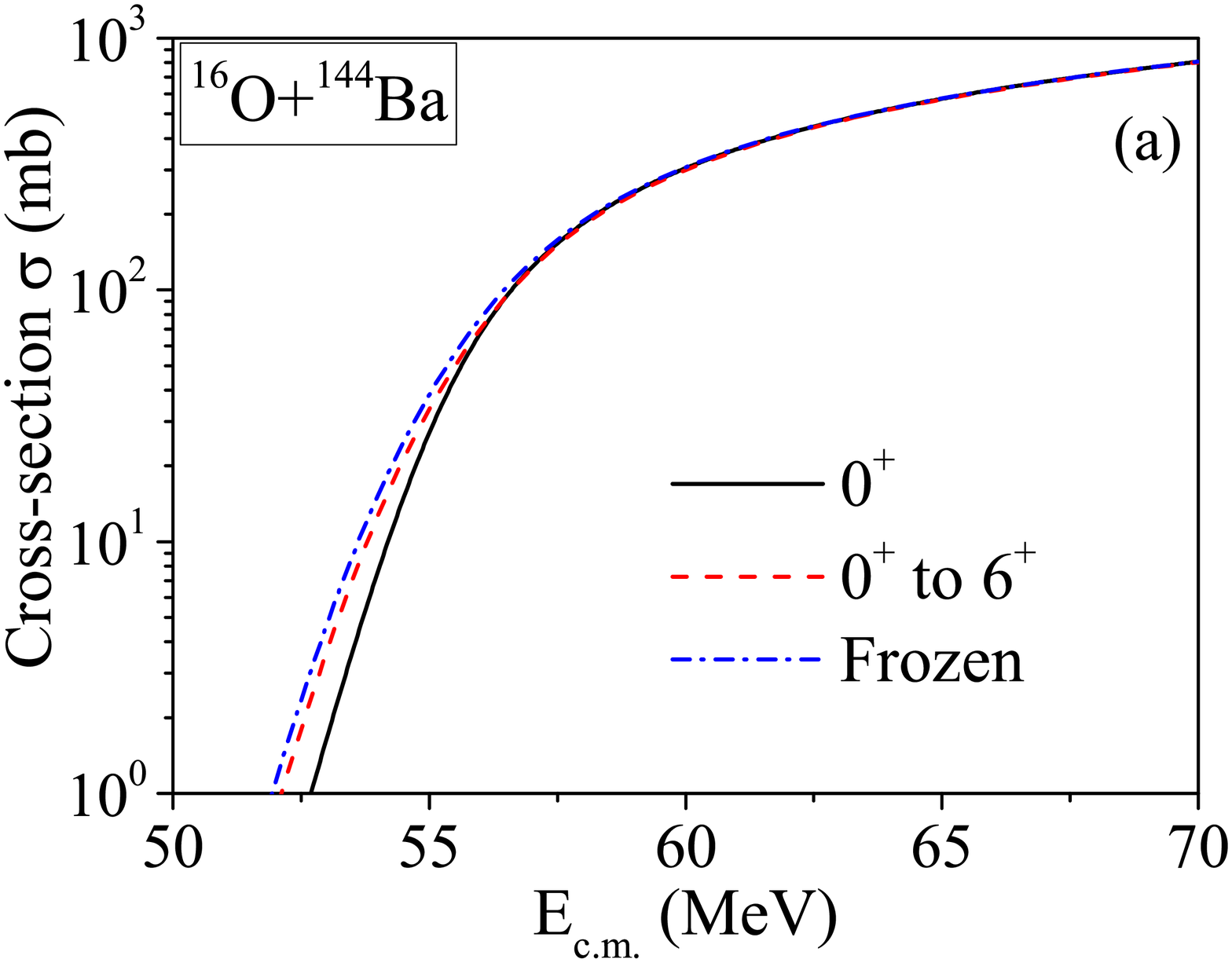}

\includegraphics[width=0.9\columnwidth]{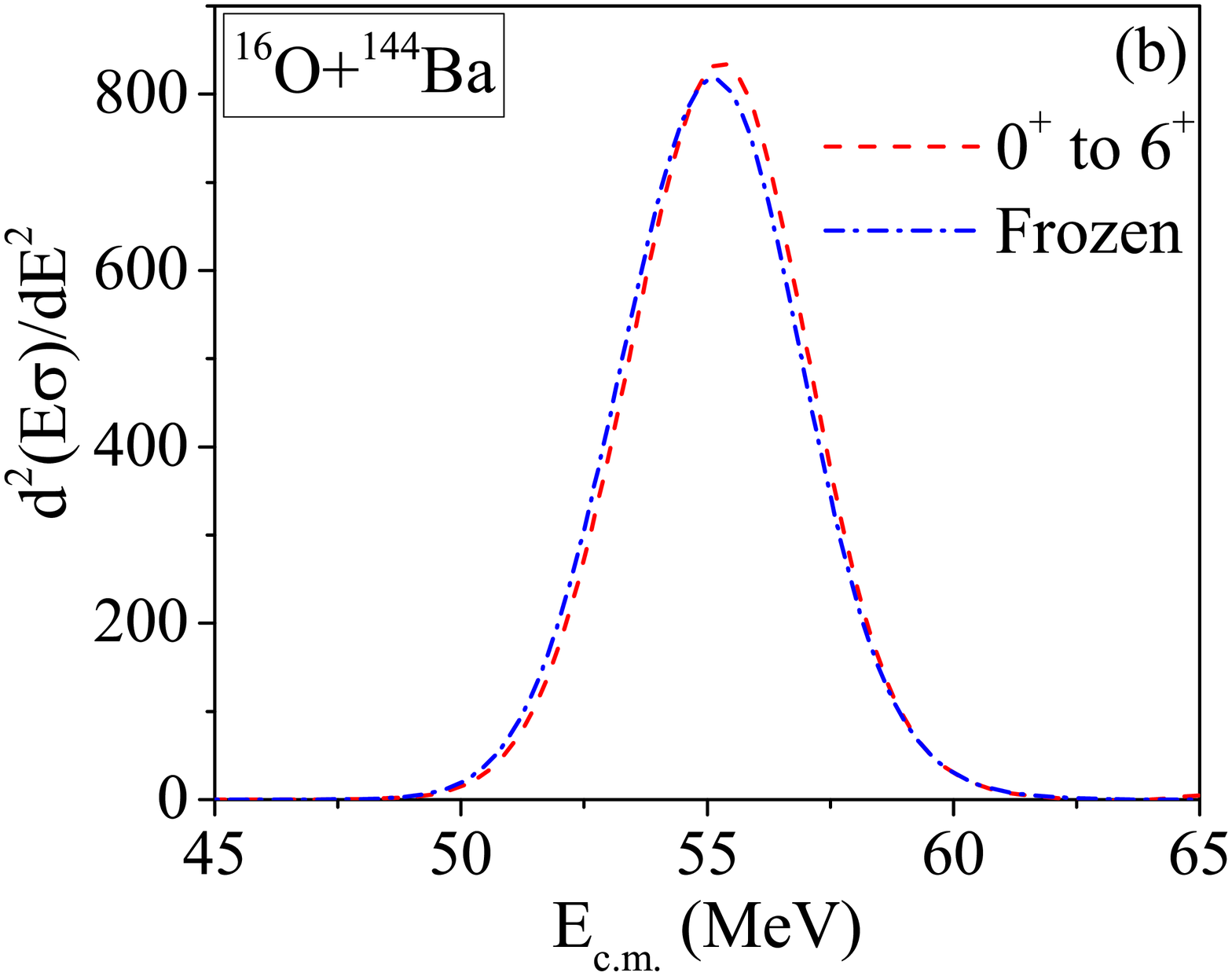} 
\vspace{-0.4cm}
\caption{(Color online) Cross sections (upper panel) for the fusion of $^{16}$O and $^{144}$Ba considering $^{144}$Ba as a spherical nuclei (solid line) and including a positive quadrupole deformation within the frozen approximation  (dashed-dotted line) and within the coupled channels framework (dashed line). For the last two, barrier distributions are shown in the lower panel.}
\label{frozen}
\end{figure}

The effect of static deformation was previously studied in terms of the frozen approximation in~\cite{Das90} for quadrupole and in~\cite{Cat89} for octupole deformation. Within this formalism the energies of the different states are neglected. As a consequence, the total cross section can be defined as the average cross section between those obtained in a single channel calculation for the different barriers found for each possible orientation of the deformed nucleus. This fact simplifies the calculations but it has been shown to overestimate the cross sections at energies below the barrier~\cite{Bal98,Esb83}. However, the first states of the nuclei investigated have small excitation energies, what should reduce the possible overestimation.


In order to test the validity of the frozen approximation for these nuclei, we have studied the fusion of $^{16}$O and $^{144}$Ba considering only the quadrupole deformation. In the Fig.~\ref{frozen} we show the result of the Coupled Channel calculation as in the previous subsection and the result of considering the frozen approximation for the same deformation parameter $\beta_{2}$. We see in Fig.~\ref{frozen}(a) that, at energies below the barrier, the frozen approximation slightly overestimates the cross section. Barrier distributions are shown in Fig.~\ref{frozen}(b). Main qualitative features are kept so that the frozen approximation is consistent with the coupled channel calculations for this case. Therefore, this aproximation can still be of great interest but should be managed with certain caution.

\subsection{Difference between octupole deformation and vibration}

\begin{figure}
\includegraphics[width=0.48\columnwidth]{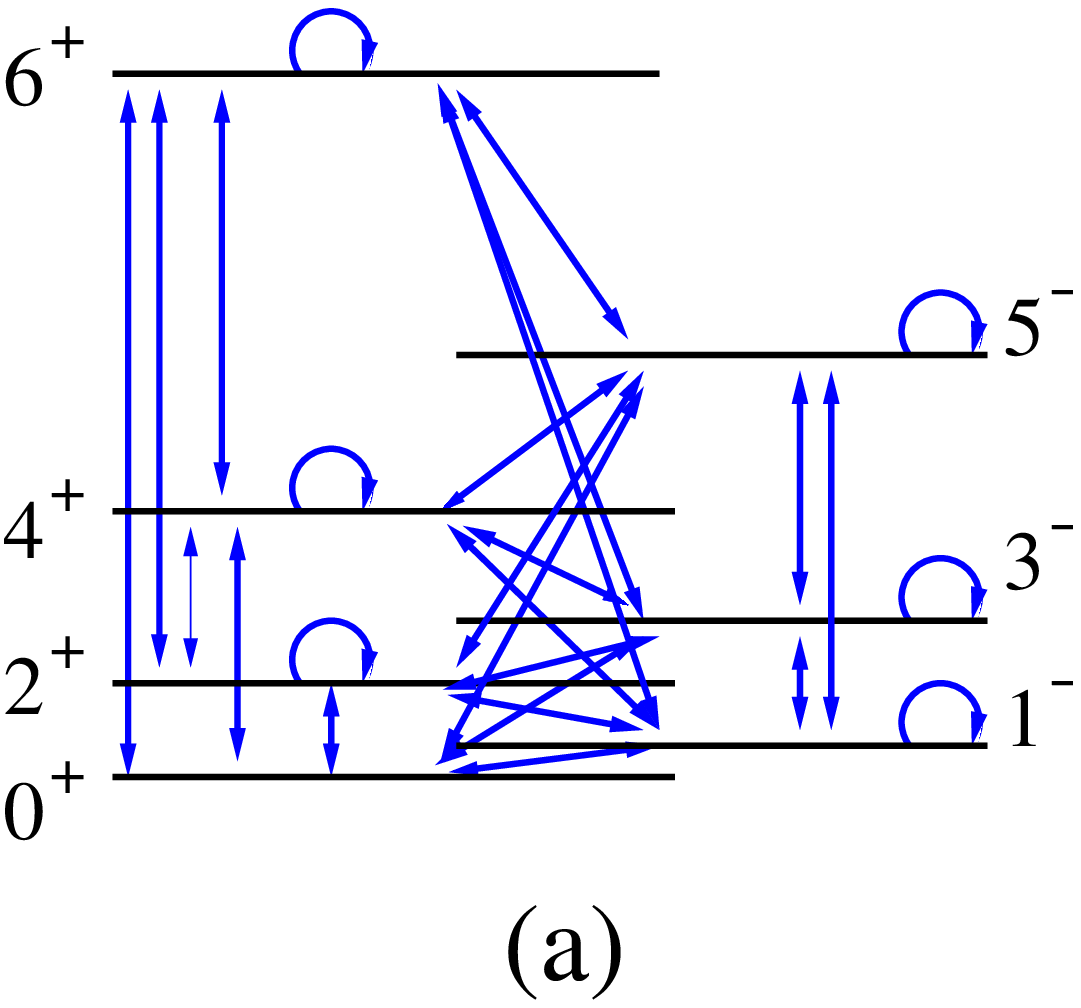}\hfill\includegraphics[width=0.48\columnwidth]{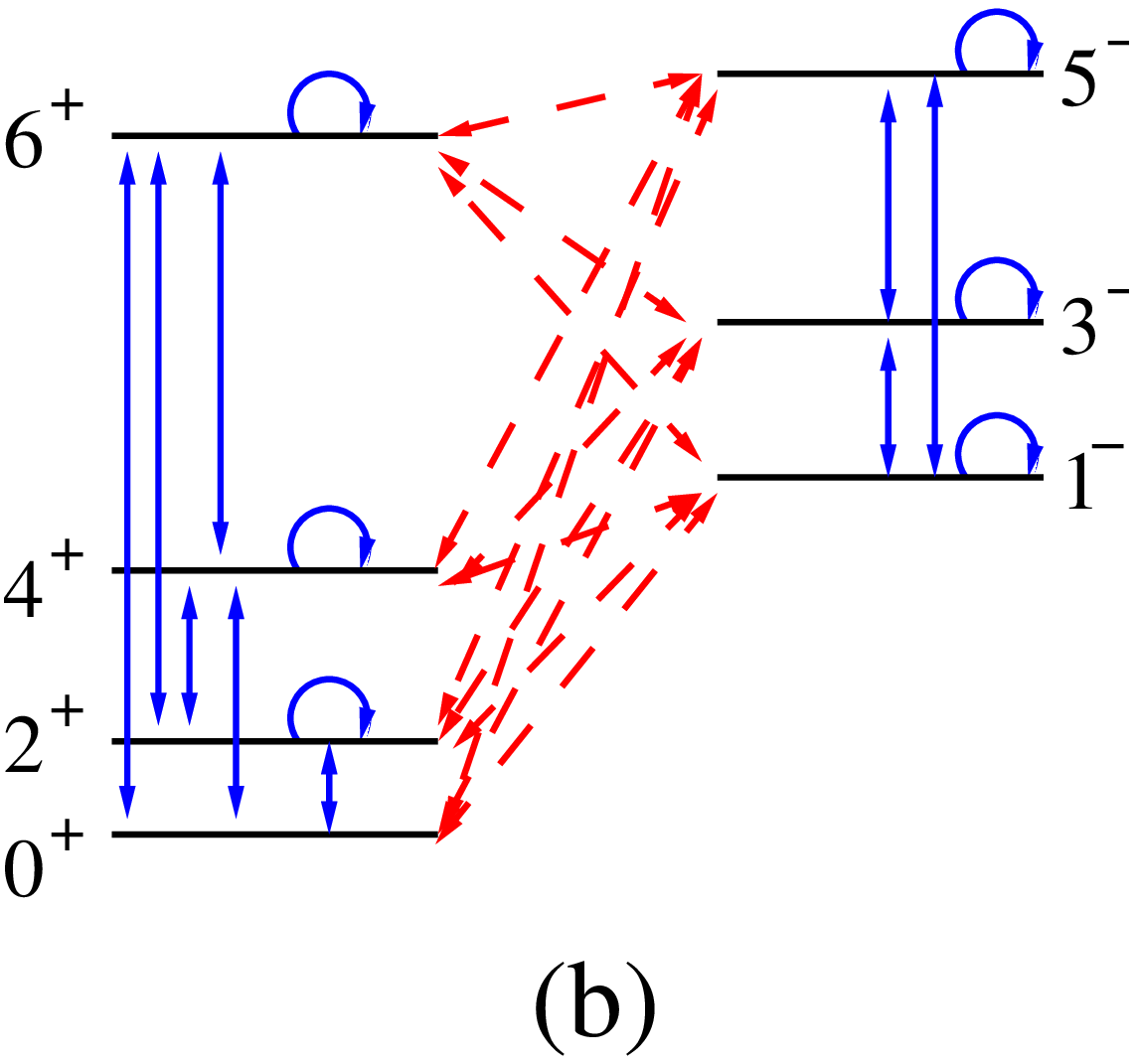}
\caption{(Color online) Level schemes considered for a quadrupole octupole deformed nucleus (left panel) and for a quadrupole deformed nuleus with an octupole vibration (right panel). Each arrow can be related to transitions with more than one possible multipolarity $\lambda$.}
\label{levelsch}
\end{figure}

\begin{figure}
\includegraphics[width=0.9\columnwidth]{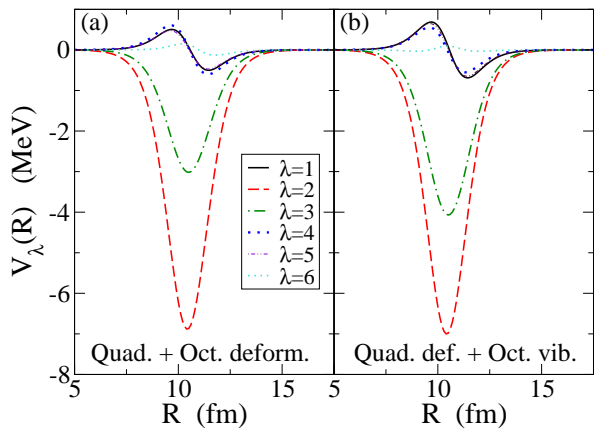}
\caption{(Color online) Coupling potentials for the system $^{16}$O+$^{144}$Ba considering quadrupole octupole deformed $^{144}$Ba (left panel) and a quadrupole deformed nuleus with an octupole vibration (right panel). 
}
\label{formfactors}
\end{figure}

\begin{figure}
\includegraphics[width=0.95\columnwidth]{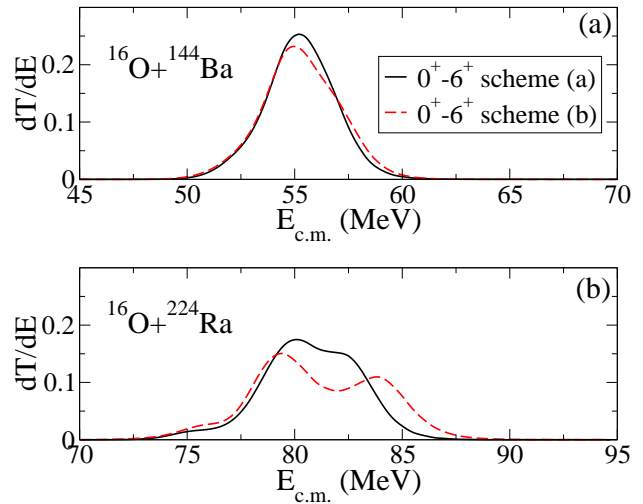}
\caption{(Color online) Barrier distribution for the fusion of $^{16}$O on $^{144}$Ba (upper panel) and $^{224}$Ra (lower panel) including a quadrupole and octupole deformation for the target (solid line) or a quadrupole deformation plus an ocutpole vibration (dashed line).}
\label{figxsrotvsvib}
\end{figure}

A more interesting and up-to-date case is to check if barrier distributions are sensitive enough to the difference between a nucleus with both quadrupole and octupole static deformations and a quadrupole deformed nucleus with an octupole vibration. The levels and the coupling potentials between the two cases will be different. In Fig.~\ref{levelsch} we show the two different coupling schemes considered. Each arrow represents the presence of at least one coupling potential with a certain multipolarity.

The scheme (a) in Fig.~\ref{levelsch} consists of a single octupole-quadrupole deformed band. Blue solid arrows in this scheme are there to remind that we can couple all this levels with a deformed potential considering the nucleus both octupole and quadrupole deformed. The scheme (b) in Fig.~\ref{levelsch} consists of a ground state quadrupole deformed rotor band and a one octupole phonon quadrupole deformed band. Different one phonon bands may result from the coupling of one octupole phonon and a quadrupole deformation. For big enough values of $\beta_2$ and $\beta_3$ the lowest energy band is the $K=0^{-}$ band with associated levels $I=1^{-},3^{-},5^{-}...$~\cite{Don66,Alo95} so that this case coincides with the octupole quadrupole deformed level scheme~\cite{BM,Eng85}.

Only red dashed arrows connect these two bands since we need a one phonon creation to go from the ground state to any of the levels in this second band.

The strength and shape of each coupling potential for the system $^{16}$O+$^{144}$Ba for each multipolarity are shown in Fig.~\ref{formfactors} again for the two different schemes considered. Differences between both schemes are small but significant. For a meaningful comparison the same value has been used for the static $\beta_3$ in case (a) and the dynamical $\beta_3$ in (b). The presence of an octupole deformation affects to the quadrupole strength and viceversa. Fig.~\ref{formfactors} only includes up to $\lambda=6$. However, larger multipolarities have been included.

\begin{table}
\caption{\label{Tab:ener} Experimental excitation energies in MeV for $^{144}$Ba and $^{224}$Ra states according to~\cite{nndc}. }
\begin{center}
\begin{tabular*}{0.45\textwidth}{@{\extracolsep{\fill}} ccccccc}
\hline
$I^{\pi}$ & $1^{-}$ & $2^{+}$ & $3^{-}$  &$4^{+}$  &$5^{-}$  & $6^{+}$  \\
\hline  
\hline  
$^{144}$Ba & $0.759$ & $0.199$  &$0.838$  & $0.530$ & $1.038$ & $0.962$ \\ 
$^{224}$Ra & $0.216$ &  $0.084$ & $0.290$&  $0.250$&  $0.433$ &  $0.479$ \\
\hline
\end{tabular*}
\end{center}
\end{table}

Even though the energies expected for the negative parity states differ in each scheme, we will use in all calculations the experimental energies~\cite{nndc} collected in Table~\ref{Tab:ener}.  Doing so, we put ourselves in the less favourable situation since the difference in energy will translate into larger differences in the barrier distribution.


With the procedure discussed in section~\ref{theo}, we calculated the barrier distributions for two systems $^{16}$O+$^{144}$Ba and $^{16}$O+$^{224}$Ra. Both  $^{144}$Ba and $^{224}$Ra are candidates to have an octupole deformation with a considerable $\beta_3$. Excitations of the projectile are not considered. Results are shown in Fig.~\ref{figxsrotvsvib}. Barrier distributions for both cases show some differences, but probably not enough to open the possibility of clearly distinguishing the two situations. It would depend on the strength of the octupole deformation and the experimental accuracy available on each case. However, it will be worthy to extend the present analysis to other regions with large octupole deformations. As just an example, $^{232}$Th could have an even larger $\beta_{3}=0.14$~\cite{But96,Sannohe} as investigated by K. Hagino and K. Sannohe~\cite{Sannohe}.

\section{\label{sum}Conclusions}

In this work we have studied the fusion reactions $^{16}$O+$^{144}$Ba and $^{16}$O+$^{224}$Ra under different assumptions for the structure of $^{144}$Ba and $^{224}$Ra. We started introducing only a quadrupole deformation and studying the sensitivity of the barrier distributions to the amount of levels included and to the possibility of having a prolate or oblate shape.

Keeping just this quadrupole deformation we have compared the results of the coupled-channel calculation with the results of the frozen approximation. We show how the frozen approximation slightly overestimates the total cross section below the Coulomb barrier. Nevertheless, the barrier distribution from the coupled channel calculation and from the frozen approximation show an overall good agreement for the present nuclei.

Finally, we analyze the effect of adding an octupole deformation or vibration. Even though the differences in the form factors are not large and the same levels are included, the fusion cross section is sensitive enough to change from one coupling scheme to the other. Both cases considered, $^{16}$O+$^{144}$Ba and $^{16}$O+$^{224}$Ra, show slight differences in the barrier distributions. This difference is larger in the $^{224}$Ra case. 
However, the final possibility of disentangle both distributions will depend on the available accuracy in a hypothetical future experiment and the strength of the octupole deformation.

Even if this difference may be not large enough for a solid conclusion towards clarifying a static octupole deformation, it is worth to say that, to our knowledge, this is the first time that octupole deformation and vibration fully coupled to a quadrupole deformation are included in a coupled-channel calculation of a fusion reaction. This should be of great help in general for the analysis of nuclear fusion with nuclei from regions with a large octupole deformation.

Since the search for a large EDM needs large static octupole deformations, it might be  possible to further ensure its presence by analyzing the subbarrier fusion cross section of some of the candidates. The possibility of finding differences in the barrier distribution could make subbarrier fusion a valuable tool to support the presence of static octupole deformations. 

To sum up, we would like to stress that fusion reactions have been of great help on understanding the structure of heavy ions for many years. In this particular case, it can also help the community to find the perfect candidate for large EDM's and, consequently, to test models beyond the standard model.

\begin{acknowledgments}
This work has been supported by MIUR research fund PRIN 2009TWL3MX. The research leading to these results has received funding from the European Commission, Seventh Framework Programme (FP7/2007-2013) under Grant Agreement nº 600376. J.A.L. is a Marie Curie Piscopia fellow at the University of Padova.
\end{acknowledgments}

\bibliography{fusiondef_rpos}

\end{document}